# Enhancing Java Call Control with Media Server Control functions

M. Femminella, *IEEE Member*, F. Giacinti, G. Reali, *IEEE Member*

DIEI, University of Perugia, Via G. Duranti 93, 06125, Perugia, Italy

email: mauro.femminella@unipg.it, francesco.giacinti@gmail.com, gianluca.reali@unipg.it

*Abstract*—In this paper, we propose a novel abstraction layer for application service implementation compliant with the Java Call Control (JCC) specifications. It simplifies creation of multimedia services using the Session Initiation Protocol (SIP) and the Media Gateway Control Protocol (MGCP). In order to show its effectiveness, we have implemented a JCC Resource Adaptor for a JAIN Service Logic Execution Environment (JSLEE), using the Mobicents application server, which is the only existing open-source JSLEE implementation. Experimental results, obtained by implementing a complex VoIP service, show both a significant simplification of service implementation and improved performance over legacy solutions.

## I. INTRODUCTION

Java is increasingly used in the development of telecom services. Java Call Control (JCC) APIs [2][3] have been introduced for alleviating service developers from the burden of handling the complexity of some communication and signaling protocols and architectures (e.g. SS7, H.323, SIP, MGCP) through the use of specific *protocol adaptors*.

The Java APIs for Integrated Networks (JAIN) Server Logic Execution Environment (SLEE) have been introduced to support developers in implementing, deploying, and managing advanced telecom services [1]. A JSLEE server allows integrating different resources and protocols in order to realize communication platforms able to fulfill the performance requirements of multimedia services, in terms of latency, throughput, consistency, and availability.

Commercial JSLEE JCC adaptors currently exist for intelligent network (IN) protocols only (e.g. CAMEL, CAP, INAP). Even the main JSLEE implementers (OpenCloud [4] and Amdocs [5]) have not provided a JCC support for SIP and/or MGCP yet, probably for the reason that SIP and MGCP are rather different from IN protocols, and the JCC-SIP mapping is not straightforward. Some proposals of mapping JCC onto SIP already exist (see [7] and references therein). Nevertheless, none of them consider JCC mapping onto MGCP. This is an important limitation since MGCP is the mainly signaling protocol currently used a for managing Voice over IP (VoIP) calls by external media gateway controllers, both for interworking with PSTN and for controlling media servers [13].

The creation of telecommunication services in a simple manner is a subject that will persist for the near future. Our contribution in this direction is using JCC as an abstraction layer on top the SIP and MGCP protocols, the former for implementing signaling and call control functions, and the latter for managing interactions with MSs, and not with any generic MGs. In particular, we show a novel JCC-SIP-MGCP mapping which builds upon our recently published JCC over SIP mapping [7]. We provide a detailed description of the JCC over MGCP implementation, along with a performance comparison with a plain SIP/MGCP-based service. Our proposal revise the semantic of a JCC method. This change is specific of the JCC-MGCP mapping, and it does not require general JCC applications to be aware of what is executed underneath the JCC adaptor. All changes and extensions are thoroughly explained in what follows, along with their implementation within the Mobicents Communication Platform [6], which is the only available open-source JAIN SLEE.

The paper is organized as follows. Section II illustrates some background concepts. Section III details the motivation underlying this work. Section IV shows the JCC-MGCP operation, the implementation of which is shown in section V, along with some experimental results. Section VI reports the lesson learned, and our concluding remarks are in Section VII.

II. BACKGROUND

*A. SIP and MGCP protocol basics.*

SIP is an application-layer signaling protocol for IP networks [13]. Sessions are established through a three-way exchange between a so-called User Agent Client (UAC) and a User Agent Server (UAS). Any session starts with an INVITE message, issued by the UAC to the UAS. One or more provisional responses (e.g. 180 Ringing) are replied back by the UAS, followed by a response sent again by the UAS to the UAC (200 OK). A final acknowledgment (ACK) sent by the UAC to the UAS completes session establishment. The set of messages including a request and the relevant responses are referred to as SIP *transaction*.

MGCP is a transactional, command-response, text-based protocol [14][15]. It is used for managing a Media Gateway (MG) from a Call Agent (CA), implementing call control functions. An MG provides media playing/recording functions, including conversion between telephone circuits and data packets. The MGCP connection model is based on *endpoints* and *connections*. Endpoints can be either physical or virtual devices and can act as both data source and destination. Examples of endpoints are announcement servers, interactive voice response (IVR) access points, packet relay devices. Connections can be either point-to-point (i.e. connecting two endpoints) or multipoint (i.e. an endpoint connected to a multipoint session).

*B. Java Call Control application programming interface*

The JCC APIs have been designed for managing communication sessions through a variety of heterogeneous networks [2][3], hiding most of complexity of underlying network protocols. It includes the entities illustrated below.

- JCC Provider: it is the interface through which an application can access the implemented JCC functions. A JCC Provider can manage a number of JCC Calls.

- JCC Call: it represents a communication between two or more parties through a dynamic collection of physical/logical entities involved in a communication relationship.

- JCC Address: it is a logical endpoint (e.g. an IP address or a directory number), having a string representation.

- JCC Connection: it is a dynamic relationship between a Call and an Address; each JCC Call is composed of a number of JCC Connections.

A detailed description of the JCC APIs and the associated finite state machine (FSM) can be found in [7].

*C. JSLEE specifications*

JSLEE is a Java open standard describing an application server (AS) architecture suitable for executing telecom services. A JSLEE AS consists of a software container that provides non-logical features necessary for executing applications. This way, it can alleviate programmers from the burden of handling low-level implementation details. *Resource Adaptors* (RAs) are fundamental JSLEE component that constitute abstract interfaces with external resources, such as protocols and databases. *Event Routers* have the task of delivering events to the appropriate service modules. A number of *facilities* are also included, which are useful to implement any *service logic*. A service logic is organized in components called Service Building Blocks (SBB), which operate asynchronously by receiving, processing, and firing events. Events received from the external world are translated into internal Java events by the relevant RA. For example, incoming SIP message are handled by the SIP RA.

Currently three main implementations of the JSLEE specification are available: Rhino, a commercial AS owned by Open Cloud [4], Amdocs Service Platform, a commercial AS owned by Amdocs [5], and the open source Mobicents JSLEE [6]. Mobicents is hosted within the JBoss container [12] and includes a JSLEE container, a Media Server (MS), a Presence Server, and a container of SIP Servlets.

III. MOTIVATIONS AND RELATED WORK

Creating telecommunication services in a simple manner is a subject increasingly important. Our proposal is to use JCC as an abstraction layer on top of SIP and MGCP. The utility of this proposal goes beyond the management of simple streaming servers, such as VideoLAN. It also allows controlling MSs having the purpose of both playing and recording contents generated by users (voice and video) and capturing DTMF tones, as it extensively happens in IVR applications. In such situations, the use of MGCP (or similar protocols, such as the MEGACO [13]) is strongly recommended, since it includes these control functions.

An alternative to JCC to implement these services is using SIP Servlets (JSR 289 [1]) together with the Media Server Control API (JSR 309 [14]). Nevertheless, some drawbacks with respect to the proposed solution exists. Some shortcomings of SIP Servlets, with respect to JSLEE, are reported in [1]. In addition, JSR 289 does not provide any specific support for other telecommunication protocols and needs of gateways for communicating with non-IP networks. Differently, JCC has been created for supporting different telecommunication protocols, such as CAMEL, CAP, and INAP [3]. Thus, our JCC implementation can be extended to provide an abstract access to any other existing signaling protocols. The second issue related to the use of JSR 289 is that it requires a deep knowledge of the SIP protocol and a service developer has first to implement and handle all the SIP logic (both successful and failed cases), and then the application making use of it, which is typically time-consuming and error prone. By using our JCC implementation, the underling management of successful and unsuccessful SIP transactions is hidden to service developers, since it is automatically handled by the RA. This feature results in both a considerable decrease of the effort needed for designing and implementing software projects and a significant reduction of implementation errors.

This consideration is valid also for JSR 309. JSR 309 defines a programming model and an object model for MS control, and provides an abstraction of commonly used application functions, such as multi party conferencing. Nevertheless, it requires more object creations and method invocations to create and handle MS connections than our JCC implementation, which implies an increased effort for application developers, as shown in section V. In addition, the JSR 309 design requires much more MS resources than our JCC implementation. For instance, to play an announcement from a MS, the JSR 309 requires the creation of two MGCP connections with two MGCP endpoints, instead of a single MGCP connection required by our JCC solution. Since our goal is reducing the implementation effort without introducing any significant performance limitation, our solution is preferable.

IV. JCC-MGCP MAPPING

Our proposal of JCC-MGCP mapping builds upon the JCC-SIP mapping shown in [7]. This extension originates from the need of handling communications with MSs, which are a specific type of MG, under the JCC abstraction layer, thus avoiding the additional work required for doing it separately. In what follows we focus on the JCC-MGCP mapping only. Figure 1 shows the relationships between the implemented JCC objects, relevant to the SIP and MGCP protocols, and the JCC-MGCP Connection Handler. FSMs are also shown, including the new JCC-SIP Connection FSM transitions introduced in [7]. The large orange arrows between objects show their relationship. These arrows are oriented from the invocating object to the invoked one, whereas the included numerical range is relevant to the number of invoked instances only. For example, the 0:N relationship from the JCC-SIP Provider to the JCC-SIP Call means that the former may handle any number between 0 and N of JCC-SIP Calls. As for the JCC-MGCP Connection FSM, the dashed black arrows and the dashed boxes represent the JCC transitions and JCC states, respectively, that we have not used.

Since our aim is using JCC on top of MGCP for managing interactions with MSs, and not with any generic MGs, we have decided of using three out of the four JCC key objects, necessary for realizing the desired JCC-MGCP mapping. They are the JCC-MGCP Provider, the JCC-MGCP Connection, and the JCC-MGCP Address. We have not implemented the JCC-MGCP Call, which is not used in services involving only MSs. Hence, the JCC-SIP Call instance is in charge of selecting the type of JCC Connection to be created (SIP or MGCP) on the basis of the target address provided in the JCC createConnection() method [2][7]. A JCC-MGCP Connection can be created *if and only if* it is associated with an existing JCC-SIP Connection, with the SDP information already exchanged between parties. As for the JCC-MGCP Provider, it acts as an MGCP listener, in charge of exchanging MGCP packets with the external world. Its main functions are:

- handling associations between JCC-MGCP Connections and the relevant MGCP transactions;
- informing the JCC-MGCP Connection Handler about the any transaction timeouts;
- creating JCC-MGCP Addresses;
- sending and receiving MGCP messages relevant to JCC-MGCP Connections.

Finally, we have implemented the JCC-MGCP Connection Handler to manage the behavior of any remote MS using the MGCP connection. It acts as a bridge between each JCC-MGCP Connection and any relevant MGCP connection (i.e. MGCP leg). The rationale for adding this object is that the JCC Connection FSM is designed for providing call control functions only, thus its transitions do not allow a complete management of remote MSs. Its main tasks are:

- storing the MGCP connection parameters (e.g. endpoint types and identifiers, announcement signal path, DigitMap, connection mode, etc.) and updating them when required by applications, by using the JCC selectRoute() method of the JCC-MGCP Connection, illustrated in details in what follows;
- managing all MGCP messages (both commands and responses) relevant to an associated MGCP connection;

- requesting both JCC-MGCP Connection and JCC-MGCP Provider to fire the appropriate JCC events and to send commands and responses relevant to previous messages, respectively;
- automatically handling MGCP connection modifications.

Each JCC-MGCP Connection Handler has an associated FSM, shown in Figure 1. FSM states are described in what follows:

- Idle: an MGCP connection has not been created yet or it has been lost;
- InConnection: this is a transient state in which a handler tries to establish a connection with an MS. The first time a handler enters this state, the associated JCC-MGCP Connection changes its JCC state to CALL_DELIVERY [2][7];
- Connected: an MGCP connection has been created and can receive requests. The first time a handler enters this state, the associated JCC-MGCP Connection changes its JCC state to CONNECTED;
- Reconnection: a handler attempts to delete the existing MGCP connection and to create a new one;
- Error: an MGCP connection has been deleted due to unexpected events (e.g. a negative answer from an MS to a MGCP Modify Connection command), but the JCC-MGCP Connection is still alive and can be used to create a new MGCP connection;
- InDisconnection: this is a transient state where a handler attempts to delete an MGCP connection with a specific MS;
- Disconnected: this is the absorbing state. When a handler enters this state, the associated JCC-MGCP Connection changes its JCC state to DISCONNECTED.

Since JCC Connection specifications do not include methods to insert and change parameters, we used the selectRoute(String) method of the JCC-MGCP Connection object to implement these functions. The resulting usage of the JCC selectRoute(String) method slightly deviates from its original semantic [2]. Nevertheless, since it allows passing a String value, it seems to be appropriate using it to pass configuration parameters to the MGCP stack. When this method is invoked, the relevant JCC-MGCP Connection Handler parses the String argument and inserts/updates the relevant value.

MGCP connection parameters are classified as follows:

- Mandatory parameters: these parameters (e.g. endpoint types or IP address of the MS) are mandatory for each MGCP connection. Thus, when an application updates the value of one of them and invokes the JCC attachMedia() method of a JCC-MGCP Connection, the underlying system automatically deletes the associated MGCP connections by sending the MGCP Delete Connection commands, and creates new MGCP connections. Application services are informed of this modification by a JCC Connection MidCall event with code "CONNECTION_CHANGED".
- Modify parameters: these parameters (e.g. SDP parameters, endpoint type) may be updated on the relevant MGCP connection through a single MGCP *Modify* command. As above, this is done automatically by the underlying system. Application services are informed of this change by a JCC Connection MidCall event with code "CONNECTION_CHANGED".
- Request parameters: these parameters (e.g. audio path, DigitMap) are relevant only to a request and are used to create the MGCP Notification Request command that instructs MS to execute a task.

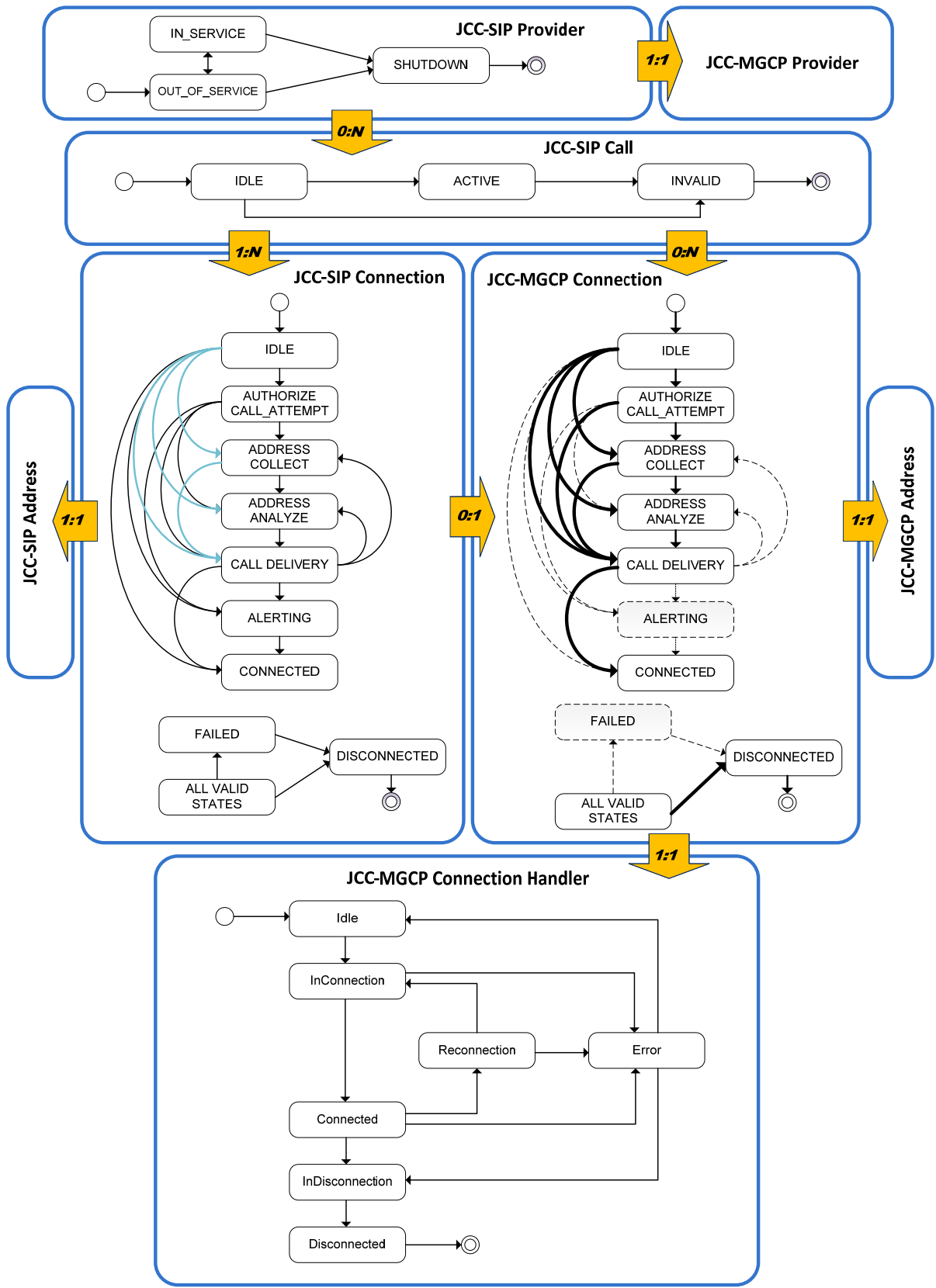

Figure 1: Relationships between the implemented JCC objects and the JCC-MGCP Connection Handler, with the relevant FSMs, including the new JCC Connection FSM transitions introduced in [7].

Finally, a JCC-MGCP Address is an object representing an MS entity. This object is created when a createConnection(A,B,-,-) JCC Call method is invoked [2], where A is an MS address (IP address and port number) with the "mgcp:" prefix, and B is the address of an existing JCC-SIP Connection. Each JCC-MGCP Address object has a one-to-one association with the relevant JCC-MGCP Connection.

In Section IV.A, we illustrate the signaling exchange of a sample service used to explain the effectiveness of our solution. This service, which implements a Prepaid Card Service (PCS) management procedure, uses both the MGCP and the SIP protocols, and thus it is a good candidate to highlight the main features of our proposal. Then, in Section IV.B, we describe the JCC-related messages in the call setup phase of the PCS service, highlighting the benefits provided by the introduction of the JCC.

*A. Description of the Prepaid Card Service*

Figure 2 illustrates the signaling exchange of the PCS service. It consists of a third party call control based on a central back-to-back UA (B2BUA). The caller (UAC) has a credit available on a prepaid phone card and desires to make a call towards a callee (UAS). The audio menu, provided by the MS, requests the caller to insert his card number, together with the PIN code, and the callee address. Then, after validating the authentication data provided by the caller through DTMF tones, the service decides whether the call request is admitted or rejected.

The control service is implemented through a Mobicents JSLEE (MSLEE), which acts as an AS implementing both the SIP B2BUA and the MGCP CA. It makes use of the proposed JCC-SIP-MGCP RA as a bridge connecting the JSLEE logic and the underlying signaling protocols. Five main signaling entities are involved: a SIP UAC, a SIP UAS, an MSLEE AS, an MS, and a Database (DB), which is used to store user profiles and service parameters.

The signaling exchange begins when the SIP UAC sends a SIP INVITE message to the MSLEE. When the MSLEE receives this INVITE message, an event routing subsystem is invoked and a Selector root SBB is created. It immediately queries the DB for obtaining the subscriber profile information. This way, the Selector SBB has all the information needed for activating a child SBB, called PrepaidCardCallControl, and leaving the signaling flow control to it. Now, the child SBB may use the SDP information provided by the UAC for creating a communication session with the MS. To do this, a Create Connection (CRCX) MGCP command is sent to the MS including the UAC SDP. The MS allocates the resources needed for this connection, generates its SDP information field, and transmits it back to the MSLEE within the subsequent 200 (CRCX) MGCP response.

The MS SDP is then forwarded to the UAC within the 183 (Session Progress) SIP response to allow it to establish a communication session between the MS and the UAC. At this time, the service logic, executed in the MSLEE, instructs the MS to play an audio message to the caller and/or indicates which UAC inputs the MS has to listen to. This is done through the Notification Request (RQNT) MGCP command. When this task is accomplished, a Notify (NTFY) MGCP command is sent from the MS back to the MSLEE. This message informs the MSLEE about both the task success or failure, and the list of information provided by the caller. After validating all information, the child SBB starts creating a second call leg towards the UAS by sending a SIP INVITE message to it, including the UAC SDP. The SIP signaling exchange proceeds until the two SIP three-way-handshakes are completed and a session is established. Periodic DB queries are employed to implement the operator-defined billing profiles and store the remaining credit within the DB during a call lifetime. Upon call termination, a final DB query updates service-related information stored in the DB. In the example shown, the call is closed by the AS by sending a BYE messages towards each of the two SIP parties, as it typically happens when a credit expires.

## B. Mapping JCC messages onto the PCS call setup

Now, we focus on the JCC-related messages in the call setup phase of the PCS, and highlight the exchanges occurring *within* the AS that controls the call evolution, as shown in Figure 3.

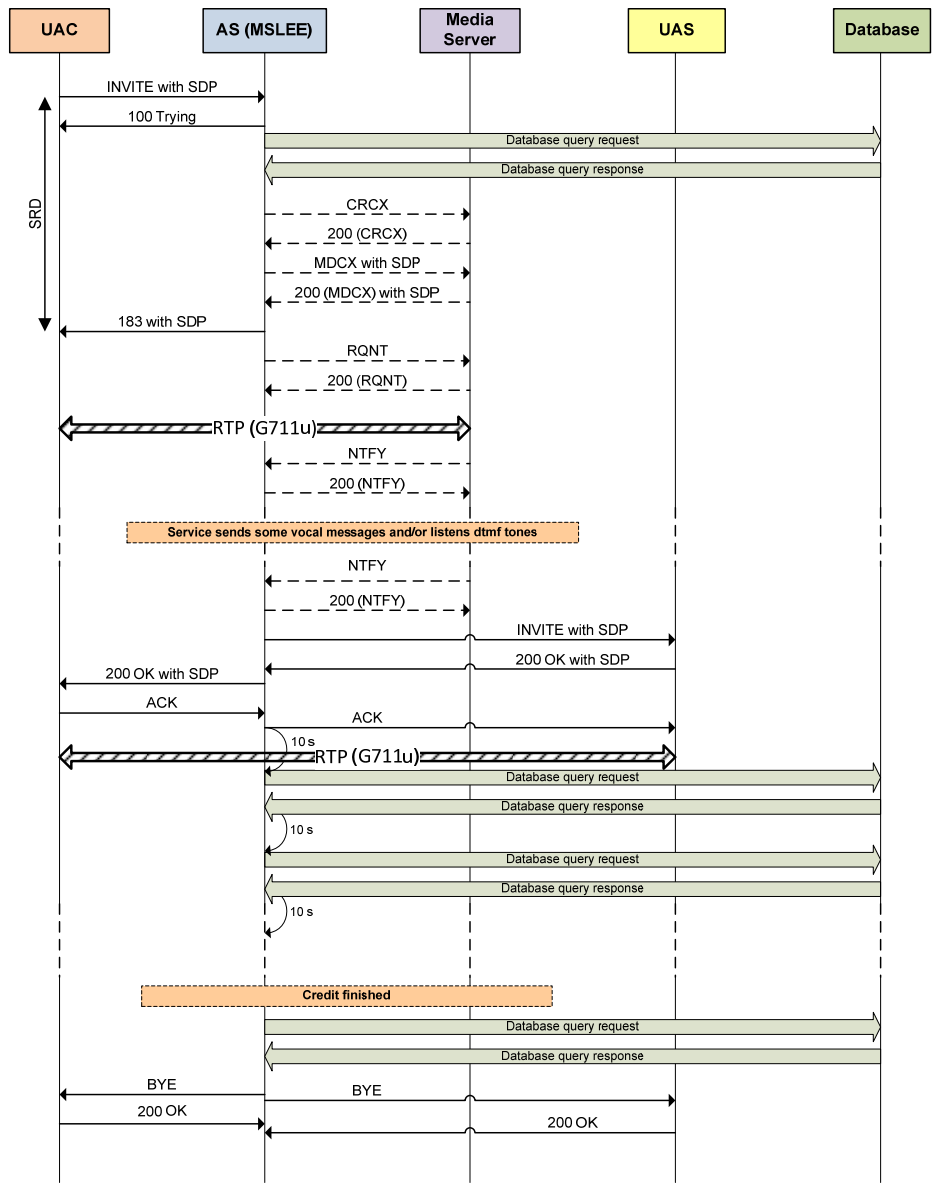

Figure 2: Prepaid card service signaling flow.

The involved entities are: JCC Call, JCC-SIP Connection A, JCC-SIP Connection B, JCC-MGCP Connection C, JCC-SIP Provider, JCC-MGCP Provider, MS, and SIP End Parties (UAC and UAS). In order to preserve readability and neatness of the figure, arguments of JCC methods are not shown. The current state of the JCC Call and Connection objects is shown in dashed boxes. The entities JCC Call, JCC-SIP Connection A and B, JCC-MGCP Connection C, JCC-SIP, and JCC-MGCP Providers are all implemented within the JCC-SIP-MGCP RA, and the methods implemented for the signaling exchange between them are RA internal methods (blue arrows). Thus, they are *completely transparent* to application developers, who have to implement only the SBB relevant to the application and manage only messages corresponding to red solid lines in Figure 3. Upon receiving an external

SIP INVITE, the steps necessary to establish a communication between an MS and a SIP UA, represented by a JCC-SIP Connection [7], are the illustrated below. We focus only on the JCC-MGCP related messages, since those relevant to JCC-SIP are extensively explained in [7].

1. The JCC Application invokes the createConnection(C,A,-,-) JCC Call method to create a new JCC-MGCP Connection, where C is the string that represents the MS address (IP and port) with the "mgcp:" prefix, and A the address of an existing JCC SIP Connection (messages 9-11 in Figure 3);

2. the JCC Application sets the endpoint type by invoking the selectRoute() method on the JCC MGCP Connection (message 12). The supported endpoint type are Announcement, IVR, Packet Relay with Announcement or Packet Relay with IVR;

3. the JCC Application invokes the routeConnection(false) method on the JCC MGCP Connection (message 13) to start the message exchange so as to both create the connection on the MS and allow UA and MS to exchange SDP information (messages 14-25). This exchange allows selecting suitable codecs for the communication between the MS to the SIP UA. Signaling message selection and exchange is automatically done by the underlying system, according to the state of the relevant JCC-SIP Connection. It is completely transparent to application developers, who have to implement only the application service logic;

4. the JCC Application waits to be notified of the transition of the JCC-MGCP Connection state to CONNECTED (message 26). This event indicates that the MS connection has been created, it is ready to receive requests, and the SDP information has been exchanged.

5. the JCC Application triggers the transmission of MGCP requests by invoking JCC methods in the following order: i) selectRoute() invocations to specify the parameters of MGCP request, one invocation for each parameter until all of them have been communicated (message 27), ii) one attachMedia() invocation on the JCC MGCP Connection (message 28). After these operations, the JCC Application waits for the request results (e.g. successful or failed, DigitMap matched or not, time interval between messages 29 and 34), which will be delivered with a JCC connection MidCall event (message 35). The above sequence of signaling messages can be repeated M times depending on the application service logic;

6. the JCC Application invokes the release() method on the JCC-MGCP Connection to delete the MS connection when either there are no further requests to be sent or the application terminates it (message 36). This is acknowledged by the JCC-MGCP Connection by invoking the connectionDisconnected() method (message 41).

7. now the JCC Application can establish a call between SIP parties by invoking the routeCall(B,A,-,-) method (message 42), and waits for acknowledgements from the relevant JCC-SIP Connections. They notify the JCC Application of this change by firing events generated by invoking the connectionConnected() method (messages 59 and 60). The call is now established and the SIP parties A and B can communicate by using the RTP protocol.

By looking at Figure 3, it is quite evident that using the JCC it is possible to reduce the number of messages to be handled by the application SBB. Two metrics are typically used to evaluate the effort in realizing software projects, the source lines of code (SLOC), and the required man-months evaluated by the *constructive cost model* (COCOMO) [11], which takes into account the SLOC value. In the next section, we will present the values of both these metrics for different implementation alternatives, in order to quantify the benefits introduced by using JCC for implementing the PCS service.

## V. PERFORMANCE EVALUATION

The PCS performance evaluation has been accomplished as follows.

First, we compare the performance of a legacy system, which uses the SIP and MGCP RAs methods available in the MSLEE, with that obtained using the JCC, in order to highlight the improvements in terms of throughput and setup time. We present two different implementations using the JCC APIs. The former one uses the methods of our JCC-SIP-MGCP RA implementation. The latter one still uses our JCC-SIP-MGCP RA, although the service logic uses the jBPM workflow engine, that we have integrated in the MSLEE to further simplify service creation [10].

Second, we evaluate the overall maximum throughput achievable using the new JCC APIs for the two implementations.

Finally, we compare the SLOC and COCOMO metric values, including in this comparison also the usage of SIP RA and MS Control RA (implementing the JSR 309 API) available in the MSLEE.

The JCC RA has been implemented by using the MSLEE v.2.7.0.FINAL, which is compliant with the JAIN SLEE 1.1 specifications. The interaction between the RA and the underlying SIP network has been realized by using the JAIN SIP API and JAIN SIP Reference Implementation (RI) v1.2, while the MGCP part has been implemented through the JAIN MGCP RI v1.0. The same RIs have been used to implement the SIP and MGCP providers in the JCC RA.

The SIP UAC and UAS have been implemented by using the SIPp traffic generator [7][8][10]. The MSLEE instances have been virtualized by using the VMWare ESXi 4.1 hypervisor, installed on a server with dual Intel Xeon E5410 (8 CPUs) and 32 GB of RAM. One virtual machine (VM) hosts the MSLEE running the PCS service. It is configured either with 2 or 4 virtual CPUs (vCPUs) and 7 GB of RAM; the JVM heap size is 6 GB. The other VMs host the MSs implemented by a Mobicents MS v3.0.0.FINAL, configured with 2 vCPUs and 5 GB of RAM; the JVM heap size is 4 GB. This configuration is motivated by the results presented in [8], which show that virtualization optimizes the usage efficiency of the computing resources for Java servers. Subscriber policies have been stored within a MySQL DB version 5.0.77. All VMs run the Ubuntu Server 64 bit v11.10 operating system, with the 1.6.0.30 64-bit JVM.

The SIPp UAC generated SIP calls according to a deterministic arrival process with a call length of 180 seconds. The duration of each test was 60 minutes. We have recorded both the maximum call throughput (MCT), expressed in calls per second (cps), defined as the maximum sustainable load with a fraction of lost calls lower than 0.01, and the session request delay (SRD), defined as the time interval from the initial SIP INVITE to the first non-100 provisional SIP response (see Figure 2). SRD values are related to the latency experienced by a caller initiating a session.

In our experiments, the most significant contribution to the SRD is the processing time within ASs. Hence, we consider acceptable an input load if the 95-th percentile of associated processing time is below 500 ms [8]. Outcomes with average SRD values larger than 1 second or call loss rate larger than 1% are not shown.

Figure 4 shows the SRD versus throughput. Figure 4.a shows the average latency with 95% confidence intervals, whereas Figure 4.b shows the 95-th percentile of SRD, both expressed in ms. Since the 95-th percentile of SRD is below the threshold of 500 ms for all implementations, we just compare the MCT. Using JCC, also combined with jBPM, allows increasing the MCT by approximately 37% with respect to the legacy implementation.

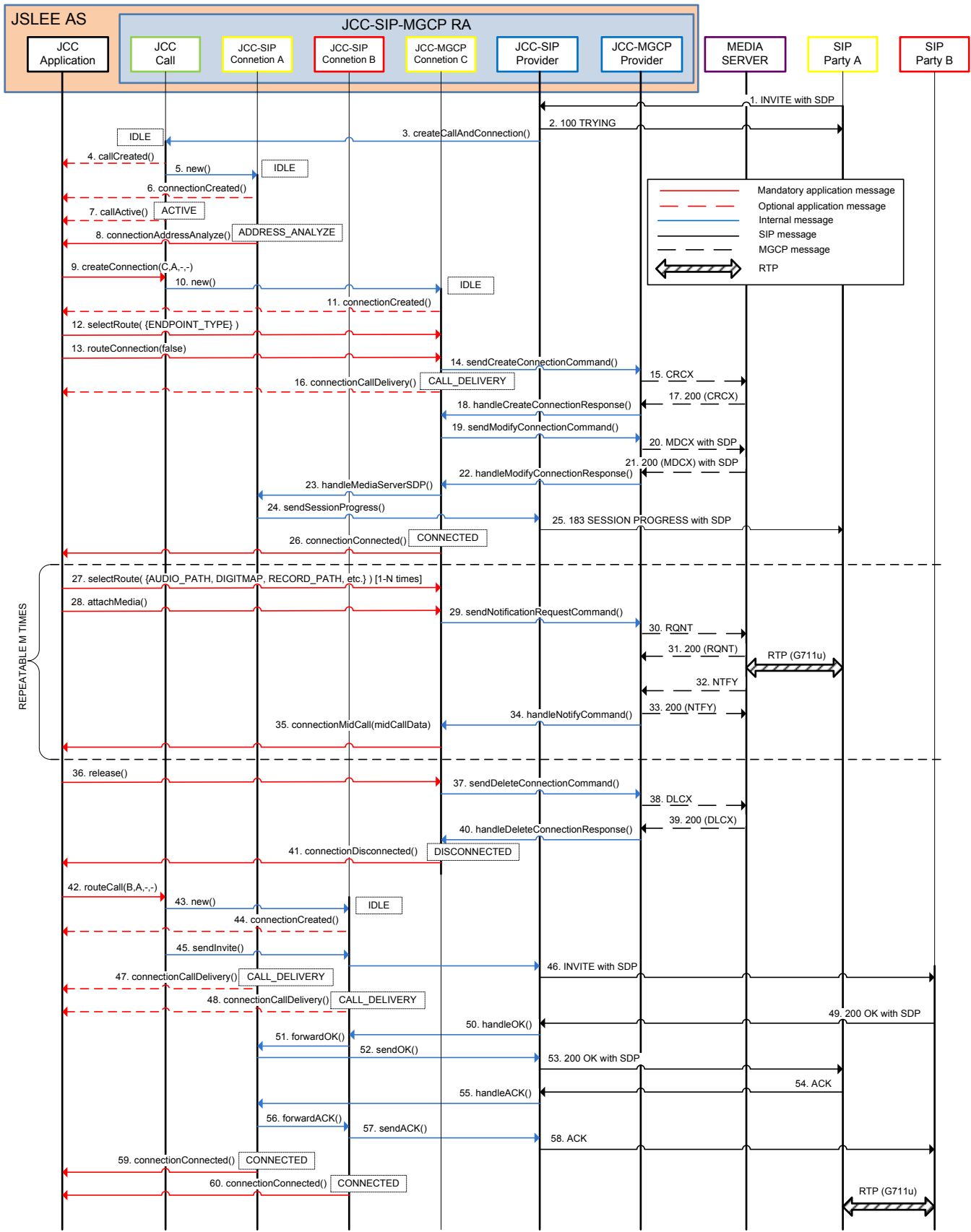

Figure 3: JCC-SIP-MGCP signaling exchange for call setup.

We have also investigated whether the system bottleneck is either the MSLEE configuration or the MS. Results are presented in Figure 5, whose abscissa and ordinate are the same of Figure 4. Each curve is relevant to a different number of both MSs and vCPUs of the MSLEE, for both JCC and JCC+jBPM approaches. An increase of the number of MSs leads to an MCT increase.

When a single MS is used, it limits the achievable performance. In fact, when the MS load approaches the maximum tolerable value, some requests cannot be processed by the request timeout and are retransmitted, thus producing a negative effects on both servers. When two MSs are used, they result quite unloaded, and their response time is always short and do not trigger MGCP requests retransmissions. In this case, the MSLEE is the performance bottleneck. Any increase of the number of vCPUs in the MSLEE can provide further improvements, due to the increased processing capabilities for serving MGCP requests. This is true also when jBPM is used. However, in this case the JCC outperforms it.

Table 1 reports a summary of the achieved results. The column labeled MCT-95 reports the MCT with a 95-th percentile of SRD lower than 500 ms. Table 1 also includes the values of SLOC and COCOMO to evaluate the effort for realizing the PCS service using the legacy solution, the SIP+MS Control (JSR 309) RAs, the JCC RA, and the JCC RA with the jBPM. The throughput of the SIP+MS Control implementation has not been evaluated, since the MS Control RA in the MSLEE is not optimized. However, it is worth to note that its performance is certainly upper bounded by the legacy solution. Hence, not only the proposed RA provides a significant reduction of the implementation effort, but it also allows improving performance.

The JCC allows saving about 30% in development effort, and exhibits the highest achievable throughput. When combined with the jBPM, it allows a further reduction of the implementation effort (-48.6%), at the expenses of a slight throughput penalty with respect to using the JCC alone, only when 2 MSs are used (-17.7% with 2 vCPUs, -23.5% with 4 vCPUs).

## VI.  LESSON LEARNED

A significant issue faced during the RA design is that the JCC API does not include methods to handle parameter values. Thus, as illustrated in section IV, we have introduced these features for supporting service implementation on top of the protocols that require adding/updating/deleting different types of parameters, such as MGCP.

Moreover, during the implementation of the JCC RA, we had to reduce the number of objects to be created in the JCC Connection FSM. In fact, the standard JCC FSM has been designed for providing call control functions only. In particular, if a JCC Connection object enters the CONNECTED state, the only possible transitions are into FAILED or DISCONNECTED states (see Figure 1). Hence, if any MS parameter values have to be changed by the service logic, and the JCC Connection object is in the CONNECTED state, it would be necessary to delete it and create a new object including the new parameter value. This approach would lead both the implemented RA and any service using it to handle a lot of new events, messages, and Java objects. This inappropriate way of proceeding has been avoided by implementing an internal JCC-MGCP Connection Handler.

A crucial aspect is that, being the MSLEE an event-driven platform, thread management within the RA has a significant impact on throughput and latency. Thus, it became necessary to optimize the thread management to reduce the number of the strictly necessary Java objects handled in the first version of the JCC-SIP-MGCP RA. Our advice for any RA developer is dedicate a significant amount of time for optimizing thread management, otherwise it is likely to incur system faults, performance degradation, and waste of resources in operation, when the system is high loaded. Our thread management optimization regards essentially the choice of the optimal number of threads for a fixed thread pool. We have found that the use of fixed thread pools is more convenient than using cached thread pools.

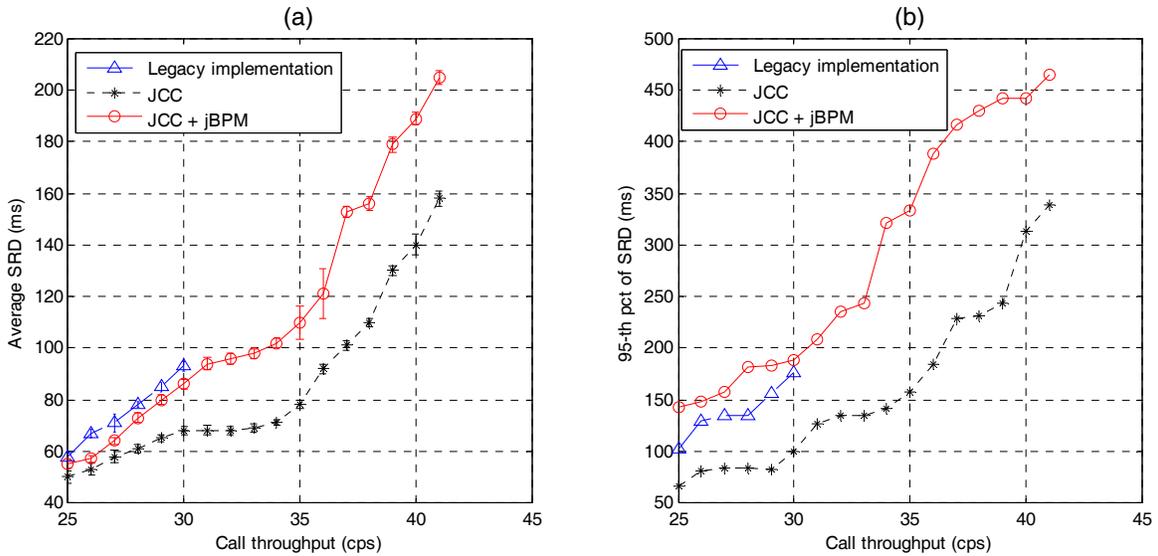

Figure 4: Setup latency vs. server throughput; single MS configuration.

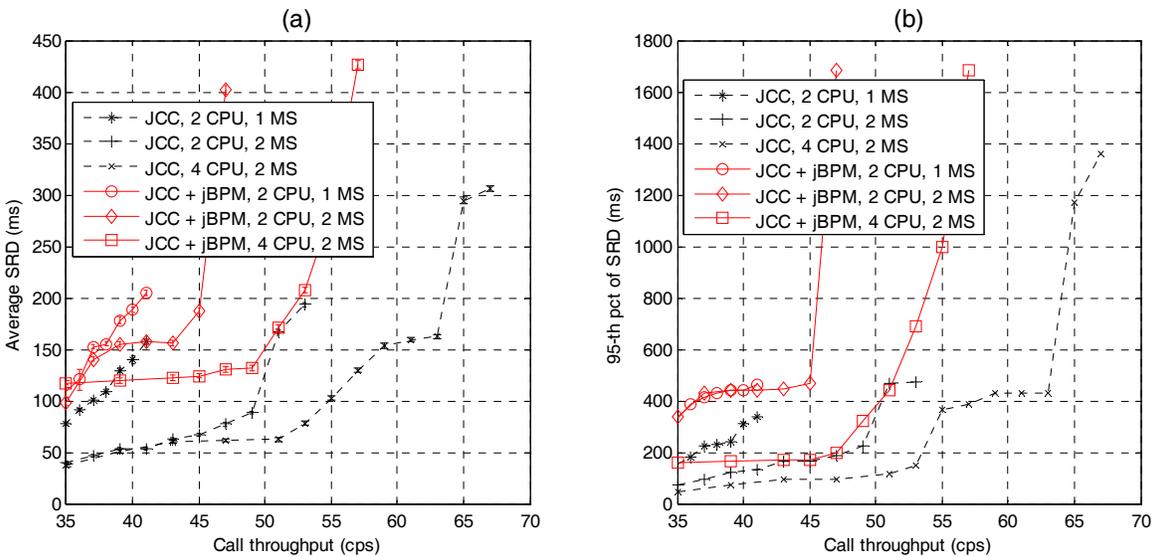

Figure 5: Setup latency vs. server throughput for multiple MSs.

A further key aspect is the selection of the JVM parameters, in particular the garbage collector (GC) type, since it can pause the running application for an unpredictable time. In order to cope with this problem, we have tuned it experimentally. In synthesis, the most suitable configuration of the implemented RA consists of using the Parallel GC with the parameter NewRatio [9] set to 10, since it is a good tradeoff between throughput and setup latency. Even if this configuration could not be optimal for all implementations, it is a convenient starting point for any experiments.

It is worth taking a critical look at the results achieved by either using additional MSs or enhancing the MSLEE configuration with additional vCPUs, reported in Table 1. Although a significant performance improvement is observed in terms achievable MCT, it could be not worthwhile to consume such an amount of resources for achieving the relevant performance improvement. Hence, although our RA allows using multiple MSs, which is also feasible by using the native Mobicents MGCP RA at the cost of introducing significant internal changes, this feature is particularly useful for handling huge traffic volumes. In this case, the best configuration consists of 3 MSLEEs for handling 4 MSs, with 2 vCPUs for each MSLEE. It is not convenient to deploy more than one MS for a single MSLEE for small traffic volumes.

Table 1 -- Summary of test configurations and results

| Approach | | Unbounded SRD | | 95-th pctile SRD ≤500 ms | | SLOC | COCOMO (mm) |
| --- | --- | --- | --- | --- | --- | --- | --- |
| *Type* | *Configuration* | *MCT* | *Avg SRD (ms)* | *MCT95* | *SRD avg (ms)* | | |
| Legacy implementation | 2 CPU, 1MS | 30 | 93 | 30 | 93 | 1333 | 3.25 |
| SIP RA + MS Control RA | 2 CPU, 1MS* | - | - | - | - | 1264 | 3.05 |
| JCC | 2CPU, 1 MS | 41 | 158 | 41 | 158 | 1049 | 2.52 |
| | 2CPU, 2 MS | 53 | 194 | 53 | 194 | 1049 | 2.52 |
| | 4CPU, 2 MS | 67 | 307 | 63 | 163 | 1049 | 2.52 |
| JCC + jBPM | 2CPU, 1 MS | 41 | 205 | 41 | 205 | 706 | 1.67 |
| | 2CPU, 2 MS | 47 | 403 | 45 | 188 | 706 | 1.67 |
| | 4CPU, 2 MS | 57 | 427 | 51 | 171 | 706 | 1.67 |

## VII. CONCLUSION

In this paper we have presented a novel JCC-MGCP mapping, which integrates an existing JCC-SIP mapping, and its implementation as a JCC resource adaptor for the Mobicents JSLEE. We have shown remarkable benefits in building complex services, such as a significant reduction of both the number of messages to be managed for establishing a media call, and the number of man-months required for implementing a service. In addition, the JCC RA implementation allows improving the achievable performance in terms of call throughputs, due to an optimized implementation of SIP and MGCP functions deployed in the RA. The integration with the jBPM tool allows reducing the implementation effort by about 49%, with a slight throughput penalty in comparison with the JCC alone.